\documentstyle[preprint,aps,graphicx,epsfig]{revtex}
\begin{document}
\font\mybb=msbm10 at 12pt
\def\bb#1{\hbox{\mybb#1}}
\def\Z {\bb{Z}}
\def\R {\bb{R}}
\def\E {\bb{E}}
\def\unit{\hbox to 3.3pt{\hskip1.3pt \vrule height 7pt width .4pt \hskip.7pt
\vrule height 7.85pt width .4pt \kern-2.4pt
\hrulefill \kern-3pt
\raise 4pt\hbox{\char'40}}}
\def\II{{\unit}}
\def\cM {{\cal{M}}}
\def\half{{\textstyle {1 \over 2}}}

\def    \beq    {\begin{equation}} \def \eeq    {\end{equation}}
\def    \bea    {\begin{eqnarray}} \def \eea    {\end{eqnarray}}
\def    \lf     {\left (} \def  \rt     {\right )}
\def    \a      {\alpha} \def   \lm     {\lambda}
\def    \D      {\Delta} \def   \r      {\rho}
\def    \th     {\theta} \def   \rg     {\sqrt{g}} \def \Slash  {\, /
\! \! \! \!}  \def      \comma  {\; , \; \;} \def       \pl
{\partial} \def         \del    {\nabla}

\preprint{UG-00-10}

\title{Flowing from a noncommutative (OM) five brane via its supergravity
dual}
\author{D. S. Berman\footnote{d.berman@phys.rug.nl} and
P. Sundell\footnote{p.sundell@phys.rug.nl} }
\address{Institute for Theoretical Physics, University of Groningen, \\
Nijenborgh 4, 9747 AG Groningen, The Netherlands }

\maketitle
\begin{abstract}

The dual supergravity description of the flow between (2,0)
five--brane theory and the
noncommutative five--brane (OM) theory is examined at critical five--brane
field strength. The self--duality of the field strength is shown to
arise as a consequence of the supergravity solution. Open
membrane solutions are examined in the background of the
five--brane giving rise to an M analogue of the noncommutative open
string (NCOS) solution.

\end{abstract}

\section{Introduction}

The space/time noncommutativity that arises on D-branes in the presence of
a near critical `electric' Neveu-Schwarz potential has produced some
interesting surprises. In particular on a D3 brane after taking a certain
limit, one is left with a new noncommutative open string theory (NCOS)
that is decoupled from closed strings \cite{GMMS,lennat}. One natural
question is to
determine the M--theory origin of these NCOS. The five--brane plays the
role of the D-brane and the open membrane plays the role of the string.
The background three form $C$ then plays the role of the NS two form. This
has motivated the investigation of the five--brane theory in the
background of
a non trivial $C$ field \cite{us1,Kawamoto:2000zt}. A decoupling limit
for the five--brane that is the M--theory origin of the NCOS limit was given in
\cite{GMSS,us2}.
This theory has near critical field strength and is believed to be
associated with an open membrane theory in six dimensions. 

Previously, the
dual supergravity descriptions of different brane theories have been
investigated in several contexts, for a review see \cite{Aharony:2000ti}.
In particular the soliton that
interpolates between two different SUSY vacua has been interpreted
as providing a description of the the flow between the corresponding
decoupled brane theories \cite{Akhmedov:1998vf}. This has recently been
discussed for the NCOS in \cite{GMSS,Harmark:2000wv}. In this paper we
will 
examine some aspects of the supergravity dual of the five--brane theory.
In particular we identify the solution to eleven dimensional
supergravity that is dual to the five--brane theory at critical
field strength and also describes the flow from the conformal (2,0) theory to
the noncommutative five brane (OM) theory. This solution has been analysed
previously in \cite{Izquierdo:1996ms,Russo:1997if,cghn,Maldacena:1999mh,oz}. 
We will also describe the appropriate critical decoupling limit for the
noncommutative five--brane (OM) theory from the
supergravity point of view as an asymptotic flow.

There is one important consideration
that need to be adressed when determing the limits that one may take on
the five--brane. The adapted field strength ${\cal H}= db+f^*_5 C$ must
obey the following nonlinear self--duality constraint \cite{Howe:1997mx},

\beq 
{\sqrt{-\det g}\over 6}\epsilon_{\mu\nu\rho\sigma\lambda\tau} {\cal
H}^{\sigma\lambda\tau}={1+K\over 2}(G^{-1})_{\mu}{}^{\lambda} {\cal
H}_{\nu\rho\lambda}\ , \label{nlsd}
\eeq        

where g is the determinant of the induced spacetime metric $g_{\mu \nu}$,
$\epsilon^{012345}=1$, the scalar $K$ and the tensor $G_{\mu\nu}$ are
given by    

\begin{eqnarray} 
\label{k} K &=&\sqrt{1+{\ell_p^6 \over 24}{\cal{H}}^2}\,
,\\ &&\cr G_{\mu\nu} &=& {1+K\over 2K}\left(g_{\mu\nu}+{\ell_p^6\over 4}
{\cal H}^2_{\mu\nu}\right)\ .  
\end{eqnarray}

This presents a small puzzle; why should
the bulk supergravity three form potential obey such a self--duality
constraint? 
Here we will analyse the five--brane supergravity solution described in
\cite{cghn} and demonstrate how this solution leads to the non--linear 
self--duality of
$C$ pulled back to the five--brane and also how one may describe the
critical field limit, crucial to the NCOS construction, from the point of
view of the five--brane SUGRA solution.

We will also examine the five--brane directly in six dimensions from
the open membrane point of view. A solution
to the open membrane equations of motion in the background of near
critical field strength is presented that is the natural lift of the
string solution given in \cite{lennat}. The properties of this solution
are in accordance with the physical picture of the critical field limit
where the tension of the membrane cancels the force excerted due to the
charged membrane boundary. This solution describes how the membrane
becomes absorbed into the five--brane worldvolume which is also in
agreement with the dual supergravity solution.
Related ideas concerning the noncommutaive five--brane (OM) theory and
NCOS can be found in \cite{Lu:2000ys}.

\section{The supergravity description}

A five--brane solution of eleven--dimensional supergravity with a finite
deformation of the spacetime three--form potential was found in
\cite{Izquierdo:1996ms,Russo:1997if,cghn}. Here we will use the notation
of \cite{cghn}.
($\mu=0,1,..,5$; $p=6,..,9,11$):

\beq
ds^2=(\Delta^2-\nu^2)^{-{1\over 6}}\left(\left({\Delta+\nu\over \Delta-\nu}
\right)^{1\over 2}dx_-^2+\left({\Delta-\nu\over \Delta+\nu}\right)^{1\over
2}dx_+^2  \right)+(\Delta^2-\nu^2)^{1\over 3}dy^2\ ,
\label{ds2}
\eeq
\beq
H_{pqrs}=\ell_p^{-3}\epsilon_{pqrst}\partial_t\Delta\ ,\qquad
\Delta=k+{R^3\over r^3}\ , \quad R\equiv N^{1\over3}\ell_p\ ,
\label{vol}
\eeq
\beq
H_{\mu\nu\rho p}=\ell_p^{-3}e_{\mu}{}^ie_{\nu}{}^je_{\rho}{}^k
F_{ijk}\partial_p\Delta \label{h4}
\eeq

where $k$ and $R$ are integration constants and
$F_{ijk}$ ($i=0,1,...,5$) is the following three--form;

\beq
F_{ijk}=(\Delta^2-\nu^2)^{-{1\over 2}}\left({(\delta_i^l+{1\over
2\nu}q_i{}^l)h_{ljk}
\over 2(\Delta+\nu)^{1\over 2}}+{(\delta_i^l-{1\over 2\nu}q_i{}^l)h_{ljk}
\over 2(\Delta-\nu)^{1\over 2}}\right)\ ,
\label{f3}
\eeq
\beq
h_{ijk}={1\over 6}\epsilon_{ijklmn}h^{lmn}\ ,\quad
q_{ij}=h_{ikl}h_j{}^{kl}\ ,\quad \nu^2={1\over 24}q^{ij}q_{ij}\ ,
\eeq

The spatial line element $dx^2_+$ and the Lorentzian line
element $dx^2_-$ are given by

\beq
dx^2_{\pm}={1\over 2}\delta_\mu^i\delta_\nu^j(\delta_{ij}\pm {1\over
2\nu}q_{ij})
dx^\mu dx^\nu\ .
\eeq

The geometry of the solution was analysed in \cite{cghn}. To avoid naked
singularities the parameter $\nu$ must be restricted to $0\leq \nu\leq k$.
There are three distinct cases.
\begin{enumerate}
\item[i)] $\nu=0$ the solution is the usual five brane metric with $AdS_7
\times S^4$ in the near horizon.
\item[ii)] $0<\nu <k$, this is a noncritical field strength deformation. The
solution interpolates between the near horison
$AdS_7\times S^4$ geometry of $N$ coinciding five--branes and flat
spacetime at $r=\infty$.
\item[iii)] $\nu=k$, this is the critical field strength deformation, it
interpolates between $AdS_7\times
S^4$ and the geometry of an array of membranes stretched in the $x_-$
direction and `smeared' in the $x_+$ direction. The line element is given
below in (\ref{am}). 
\end{enumerate}
The precise justification for relating $\nu$ to the field strength on the
five--brane is given below when we analyse the properties of $C_3$. It is
the third case we wish to study. In the
asymptotic region, $r \rightarrow \infty$ one recovers the
`smeared' membrane metric:
\beq
ds^2=\left({r\over R}\right)^2dx_-^2+{R\over r}(dx_+^2+dy^2)\ ,
\label{am}
\eeq

As disussed in \cite{cghn} the solution has $16$
unbroken supersymmetries. As we flow to AdS, $r \rightarrow 0$ we have the
usual
$16\rightarrow 32$ symmetry restoration.  This metric (\ref{am}) had also
been
investigated in the context of seven dimensional domain wall supergravity
with 16 unbroken supersymmetries \cite{Boonstra:1999mp}. The fact it is
the
smeared membrane metric that appears in the world volume of the
five--brane is consistent with the OM interpretation of the critical field
limit \cite{GMSS}, see the discussion below.

We next determine the three form potential that is induced on the
five--brane worldvolume. This essentially
means that one must solve the field strength $H_{p \mu \nu \rho}$ for the
potential $C_{\mu \nu \rho}$, where

\beq
H_{\mu \nu \rho p} = -\partial_p C_{\mu\nu\rho}+3\partial_{[\mu}
C_{\nu\rho]p} \, . 
\eeq

First, making use of the algebraic properties of $h_{ijk}$ we determine,

\beq
H_{\mu \nu \rho p}=\ell_p^{-3}(\Delta^2-\nu^2)^{-2}\delta_\mu^i\delta_\nu^j\delta
_\rho^j\left((\Delta^2+\nu^2)\delta_i^l-\Delta q_i{}^l\right)h_{ljk}
\partial_p\Delta\ .
\eeq

Then, fixing a gauge so that $C_{\mu \nu p} =0$ (which preserves the
background symmetry) we may solve for $C_{\mu \nu \rho}$ as follows

\beq
C_{\mu\nu\rho}=\ell_p^{-3}\delta_\mu^i\delta_\nu^j\delta_\rho^k\left(
{(\delta_i^l+{1\over 2\nu}q_i{}^l)\over 2(\Delta+\nu)}
+{(\delta_i^l-{1\over 2\nu}q_i{}^l)\over 2(\Delta-\nu)}\right)h_{ljk}\ .
\eeq

Introducing an ${SO(5,1) \over {SO(3)\times
SO(2,1)}}$ valued sechsbein, $\{v_\mu{}^\alpha,u_\mu{}^a \}$,
$\alpha=0,1,2$,
$a=3,4,5$, of the induced metric, $g_{\mu \nu}$
at $r=\infty$ as follows:

\beq
g_{\mu\nu}=\partial_\mu X^M \partial_\nu X^N g_{MN}|_{r=\infty}
=\eta_{\alpha\beta}v_\mu{}^\alpha v_\nu{}^\beta +
\delta_{ab}u_\mu{}^a u_\nu{}^b\ ,
\eeq

we can write the pull--back of $C$ at infinity

\begin{eqnarray} 
C_{\mu\nu\rho}|_{r=\infty}&=&\ell_p^{-3}({2\nu\over k+\nu})^{1\over
2} \epsilon_{\alpha\beta\gamma} v^{\alpha}_{\mu} v^{\beta}_{\nu}
v^{\gamma}_{\rho}
+\ell_p^{-3}({2\nu\over k-\nu})^{1\over 2} \epsilon_{abc} u^{a}_{\mu}
u^{b}_{\nu} u^{c}_{\rho} \\
&=& {h\over \sqrt{1+h^2\ell_p^6}} \epsilon_{\alpha \beta \gamma}
v^{\alpha}_{\mu} v^{\beta}_{\nu} v_{\rho}^{\gamma}
+ h \epsilon_{abc} u^{a}_{\mu} u^{b}_{\nu} u^{c}_{\rho}\, ,
\qquad h^2\ell_p^6= {2\nu\over k-\nu}\ .
\label{c3}
\end{eqnarray}

The purpose of this rewriting is that we can then identify the three-form
(\ref{c3}) with the solution to the five--brane non--linear self--duality equation (\ref{nlsd}), described in
\cite{us1}. Thus the vacuum at $r=\infty$ has a non--trivial
three--form potential (\ref{c3}) in the five--brane directions, and
moreover it in fact obeys the five--brane field equation (\ref{nlsd}).
Hence we may identify

\beq
{\cal H}_{\mu\nu\rho}=C_{\mu\nu\rho}|_{r=\infty}\ .\label{fs}
\eeq

The critical field strength limit described in \cite{GMSS,us2} is when
$h^2 \ell_p^6 \rightarrow \infty$. From (\ref{c3}) we see this occurs in
the asymptotic region when $\nu \rightarrow k$. Hence this justifies our
identification of the $\nu=k$ case with the critical field strength limit. 
It is important notice that the non--linear nature of the five--brane
self--duality is crucial for the critical limit. (Linear self--duality
would not allow a critical limit.)

We now wish to describe a decoupling limit, i.e. a limit whereby
$\ell_p \rightarrow 0$, which has an asymptotic region with
five--brane metric and field strength scaling as described in \cite{GMSS,us2}.
The decoupling limit must obey the criteria that the line element
$\ell_p^{-2}ds^2$ and the four--form field
strength $H$ are held fixed (such that the supergravity action is finite
for this background; by making $N$ large the curvature becomes small and
the supergravity approximation makes sense).

Thus we are drawn to consider the following limit:

\beq
\nu=k={\ell_g^{3n}\over 2\ell_p^{3n}}\ ,\quad
\tilde{x}_\pm=\ell_g^{-{n\over
2}}\ell_p^{{n\over 2}-1}x_\pm\ ,\quad \tilde{r}={\ell_g^n r \over \ell_p
^{1+n}N^{1 \over3} }\ , \label{tilde}
\eeq

where $\tilde{x}_\pm$, $\tilde{r}$ and the length scale $\ell_g$ are
fixed and \footnote{This condition follows from considering the graviton
absorption cross section \cite{mohsen}; this was pointed out to us by M.
Alishahiha.} $n
> 0$. After rewriting in terms of fixed
variables one has:

\beq
{ds^2 \over \ell_p^2} = f^{1\over 3}\tilde{r}^2 d\tilde{x}_-^2 +
f^{-{2\over 3}} \tilde{r}^{-1}d\tilde{x}_+^2 +N^{2\over 3} f^{1\over 3}
\tilde{r}^{-1} (d\tilde{r}^2 + \tilde{r}^2 d\Omega^2)     \ , \label{dsf} 
\eeq
\beq
 H_4=N\epsilon_4(S^4)+ {1\over 2}d\left(\tilde{r}^3 d\tilde{x}_-^3 +
f^{-1}d\tilde{x}_+^3\right)\ , \eeq   
\beq 
f=1+\tilde{r}^{-3}\ . 
\eeq

This does not describe a field theory limit in the asymptotic region. It
would be wrong to expect a
field theory description of the noncommutive five--brane, just as one does
not have a field theory description for the D3 brane with temporal
noncommutativity. The five--brane with non-trivial C in a decoupling
limit has also been considered in \cite{oz} and indeed the line
element (\ref{dsf}) was found. This predated the discussion of the
noncommutative five--brane (OM) theory and the importance of the
critical field limit.

\begin{figure}[h] \begin{center}
\includegraphics[angle=0,width=140mm]{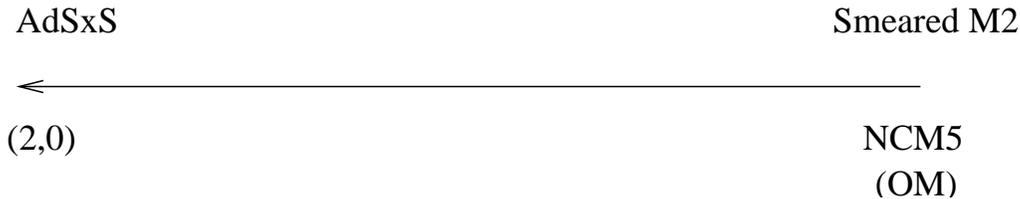} \end{center}
\caption{\small The flow as controlled by $\tilde{r}$ to the (2,0) 
theory from the NC five--brane (OM) theory is given by the interpolation
between
$AdS_7 \times S^4$ and the smeared membrane metric.} \label{flow}
\end{figure}

One may then check that scaling $\tilde{r} \sim \epsilon^{- 
{1 \over 3}}$ 
exactly reproduces the scaling taken for the noncommutative five--brane
(OM) theory \cite{GMSS,us2}:
\beq
\ell_p \rightarrow \epsilon^{ 1 \over 3}\, ,\quad 
{\cal{H}}_{\alpha \beta \gamma} \rightarrow \epsilon^{-1}\, , \quad
{\cal{H}}_{a b c}
 \rightarrow \epsilon^0\, , \quad g_{\alpha \beta}\rightarrow \epsilon^0
\, , 
\quad g_{ab} \rightarrow \epsilon ^1 \, . 
\eeq

One important property of this limit is that the open membrane metric
described in \cite{us2} is fixed in units of $\ell_p$.

In summary, the solution (\ref{ds2})-(\ref{vol}) interpolates between a
stack of five--branes at $r=0$ with zero field strength and a stack of
five--branes at $r=\infty$ with non--vanishing field strength ${\cal
H}_{\mu\nu\rho}$ given by (\ref{fs}). The solution is fixed in the limit
given by $\ell_p\rightarrow 0$ and (\ref{tilde}), which flows to
a decoupled six--dimensional noncommutative five--brane (OM) theory.

It is worth remarking that such a limit is only possible at critical field
strength. In the non-critical case, $0<\nu<k$, one cannot obtain a brane
decoupled from the bulk theory. This is consistent with how the NCOS limit
requires critical field strength to decouple the bulk modes.

\section{Critical Open Membrane Solution}

We may interpret the source of the
constant field strength at $r=\infty$ as an array of self--dual strings
stretched along the boundary of the space in the $x_-$ direction and smeared
homogeneously in the $x_+$--directions. These self--dual strings are
boundaries of open membranes. In the critical
limit, given by $\nu=1$, the open membranes become dissolved
into the five--brane. This is exactly analogous to how open strings
behave in D--3 brane with the critical field strength \cite{GMMS,lennat}.

We will now illustrate this by analysing the solutions of the membrane
field equations in the
background provided by a five--brane with critical field strength.

The membrane is described by the following equations of motion,
constraints
and boundary conditions \cite{deWit:1988ig}.

\beq
\ddot{X}^{\mu}+{1\over \ell_p^2}\{X^\nu,\{X^\mu,X_\nu\}\}=0\ ,
\eeq
\begin{eqnarray}
\dot{X}^2&=&-{1\over 2\ell_p^2}\{X^\mu,X^\nu\}\{X_\mu,X_\nu\}\ ,\\
\dot{X}^\mu \partial_i X_\mu&=&0\ ,\qquad i=\rho,\sigma\ ,
\end{eqnarray}
\beq
{1\over\ell_p^2}\sqrt{-\det{\gamma}}n^{\alpha}\partial_{\alpha}X_\mu+
\epsilon^{\alpha\beta\gamma}{\cal H}_{\mu\nu\rho}n_\alpha 
\partial_{\alpha} X^\nu \partial_{\gamma}X^\rho=0\ ,
\eeq

where $\dot{X}^\mu$ and $\partial_i X^\mu$ denote differentiation with
respect to the worldvolume time $\tau$ and spatial coordinates 
$\rho,\sigma$, respectively, 

\beq
\{A,B\}=\epsilon^{ij}\partial_i A \partial_j B\ ,
\eeq

and the determinant of the worldvolume metric is given by

\beq
\sqrt{-\det{\gamma}}={1\over 2\ell_p^4}\{X^\mu,X^\nu\}\{X_\mu,X_\nu\}\ .
\eeq

In what follows we shall choose the normal derivative at the boundary to
be given by $n^\alpha\partial_\alpha=\partial_\rho$.

In order to find a solution we make an ansatz where the open membrane
is infinitely stretched in the $X^2$ direction as follows

\beq
X^0=X^0(\tau,\rho)\ ,\qquad X^1=X^1(\tau,\rho)\ ,\qquad X^2=\ell_p \sigma
\ ,
\eeq
\beq
X^\mu=\rm{constant}\ ,\qquad \mu=3,...,9,11\ .
\eeq 

This results in the equations:

\beq
\ddot{X}^\mu-X^{\prime\prime \mu}=0\ ,
\eeq
\bea
(\dot{X}^0)^2+(X^{\prime 0})^2 &=& (\dot{X}^1)^2+(X^{\prime 1})^2\ ,\\
\dot{X}^0X^{\prime 0}=\dot{X}^1 X^{\prime 1}\ ,
\eea
\beq
X^{\prime \mu}=H\dot{X}^\mu\ ,\qquad H\equiv {h\ell_p^3\over 
\sqrt{1 + h^2\ell_p^6}}\ .
\eeq

These equations are equivalent to the equations of an open string in an
electric field $H=\alpha^\prime {\cal F}^0{}_1=\alpha'g^{00}
{\cal F}_{01}$.  

In case $0\leq H<1$ the general solution is given by

\beq
X^0=\pm X^1= a\ell_g(\tau\pm H\rho)\ ,
\eeq

where $\ell_g$ is a fixed length and $a$ a real constant.
In the case $H=1$, however, there are new critical solutions appearing.
The solution for $H=1$ is given by

\beq
X^0=\ell_g(a\tau+b\rho)\ ,\qquad X^1=\ell_g(b\tau+a \rho)\ .
\eeq

Thus we have a static, non-degenerate open membrane solution
given by

\beq
X^0=\ell_g\tau,\qquad X^1=\ell_g \rho\ ,\qquad X^2=\ell_g\sigma\ ,
\eeq
\beq
X^{\mu}=constant \ ,\qquad \mu=3,...,9,11\ .
\eeq

Notice that in the limit $\ell_p\rightarrow 0$ we have carried out a
reparametrisation of $\sigma\rightarrow {\ell_g \over \ell_p} \sigma$. 
Clearly this does not affect the geometry of the solution.

The length scale $\ell_g$ for this solution is the scale introduced in
\cite{GMSS,us2}. This is the effective tension of the open membrane inside
the five--brane and is related to the five--brane field strength by
$\ell_g=(h^2 \ell_p^9)^{1 \over3}$.
 
Thus the solution describes a membrane stretched out inside the
five--brane in
the $0,1,2$ directions. This is what one expects from the dissolved
membrane interpretation of the five--brane
supergravity solution given by (\ref{am}).

Physically, one sees that the membrane tension (which is scaled to
$\infty$) is cancelled by the scaling of the electric field so that the
membrane retains a finite length scale. This is reminiscent of the zero
force condition experienced by BPS solutions. It would be
interesting to see whether the cancellation properties persist (as they do
for BPS solutions) beyond the classical level.

An obvious application would be to investigate the thermal properties of
the noncommutative (OM) theory by analysing the non extremal version of
the smeared membrane metric (\ref{am}).

\section{Acknowledgements}

We are indebted to Mohsen Alishahiha for extremely valuable comments on
the graviton absorption cross section and the decoupling
limit.
We are greatful to Eric Bergshoeff, Martin Cederwall, Jan-Pieter van de
Schaar, Ergin Sezgin
and Paul Townsend for discussions. P.S. is grateful to  Henric Larsson and
Bengt Nilsson for discussions. D.S.B is supported by the
European Commission TMR program ERBFMRX-CT96-0045 associated to the
University of Utrecht. The work of P.S.~is part of the research program of
the ``Stichting voor Fundamenteel Onderzoek der Materie'' (FOM).

\end{document}